\documentclass{PoS}

\usepackage {amsmath}
\usepackage[export] {adjustbox}
\usepackage[caption=false]{subfig}

\title{Gamma-ray signal from Dark Matter Annihilation mediated by mixing scalar mediators}

\ShortTitle{Gamma-ray signal from Dark Matter Annihilation mediated by mixing scalar mediators}

\author{\speaker{Fei Teng}\thanks{Work done in collaboration with Jason Kumar, Kuver Sinha, Pearl Sandick and Takahiro Yamamoto}\\
        Department of Physics and Astronomy \\ University of Utah, Salt Lake City, UT 84112 \\
        E-mail: \email{fei.teng@utah.edu}}

\abstract{We present here a study of the direct and indirect detection prospects of a generic dark matter simplified model, in which the Majorana dark matter interacts only with a Standard Model lepton and a pair of uncolored mixing scalar mediators. We first show that the mixing angle significantly changes the feature of internal bremsstrahlung, as well as the flux ratio of $\gamma\gamma$ and $\gamma Z$ line signals. The $CP$-violation phase will introduce an polarization asymmetry in the $\gamma\gamma$ final state. Then we study the direct detection prospect and discuss the complimentarity of these two search strategies.
}

\FullConference{38th International Conference on High Energy Physics\\
		3-10 August 2016\\
		Chicago, USA}

\begin{document}

\section{Introduction}
The existence of dark matter {in our universe} has been a well-established fact from various astrophysical and cosmological evidence. Among all the hypotheses for the nature of dark matter, weakly interacting massive particle (WIMP) is a very attractive one, since it can reproduce the correct thermal relic density and fit in the minimal supersymmetric Standard Model (MSSM) that possibly solves the hierarchy problem of the Standard Model. Under this paradigm, dark matter, as the lightest supersymmetric particle (LSP), interacts with the visible sector by exchanging a mediator. Some well-studied examples of the mediators include squark, slepton, Higgs, chargino, etc. The fact that no new physics has been found at LHC has ramped up the lower bound to $\sim1.5\,\textrm{TeV}$ for many of these superpartners~\cite{ATLAS:2016kts,CMS:2016mwj}. On the other hand, LHC usually puts stringent bounds on those superpartners that couple to quarks, but the bounds on QCD-neutral scalars, like sleptons, are much weaker due to the messy hadronic background~\cite{Aad:2014vma,Khachatryan:2014qwa}.


In the wake of the new LHC data, we propose a simplified model in which the Majorana dark matter $X$ couples to a charged lepton $f$ through a pair of uncolored mixing scalars $\widetilde{f}_{L,R}$~\cite{Kumar:2016cum}:
\begin{equation}
  \label{eq:lag}
  \mathcal{L}_{\text{int}}=\lambda_{L}e^{i\varphi/2}\widetilde{f}_{L}^{\ast}\overline{X}P_{L}f+\lambda_{R}e^{-i\varphi/2}\widetilde{f}_{R}^{\ast}\overline{X}P_{R}f+\text{c.c.}\,.
\end{equation}
If $\lambda_{L,R}=\sqrt{2}gY_{L,R}$, our model fits in the MSSM framework with $X$ a binolike LSP. The mass eigenstates of the mediators, denoted as $\widetilde{f}_1$ and $\widetilde{f}_2$, are obtained from $\widetilde{f}_{L}$ and $\widetilde{f}_{R}$ through a rotation of the mixing angle $\alpha$. For the most generic case, our model contains seven parameters: the dark matter mass $m_{X}$, the mass eigenvalues of the scalar mediators $m_{1,2}$, the coupling constants $\lambda_{L,R}$, the scalar mixing angle $\alpha$ and $CP$-violation phase $\varphi$. The mass of $X$ is about ${O}(100\,\textrm{GeV})$, allowed by the LEP limit. We further assume that the lighter scalar mediator, say $\widetilde{f}_1$, is only $1\%$ to $20\%$ heavier than $X$, namely, it essentially lies in the ``blind spot'' of LHC. 

Using simplified models is a good strategy in face of the explosion of data. It has more predictivity than an effective field theory, while contains only those relevant degrees of freedom that can produce certain signals. Moreover, simplified model also leaves open what its UV-completion is. For example, our model \eqref{eq:lag} could be a part of supersymmetry, but it could also be a part of a very different high energy theory.

Our main purpose of this work is to study the direct and indirect detection prospects of this simplified model, and how the $\alpha$ and $\varphi$ alters the prediction.

\section{Result and analysis}
\begin{figure}[t]
  \centering
  \subfloat[IB spectrum]{%
    \includegraphics[width=0.47\textwidth,valign=m]{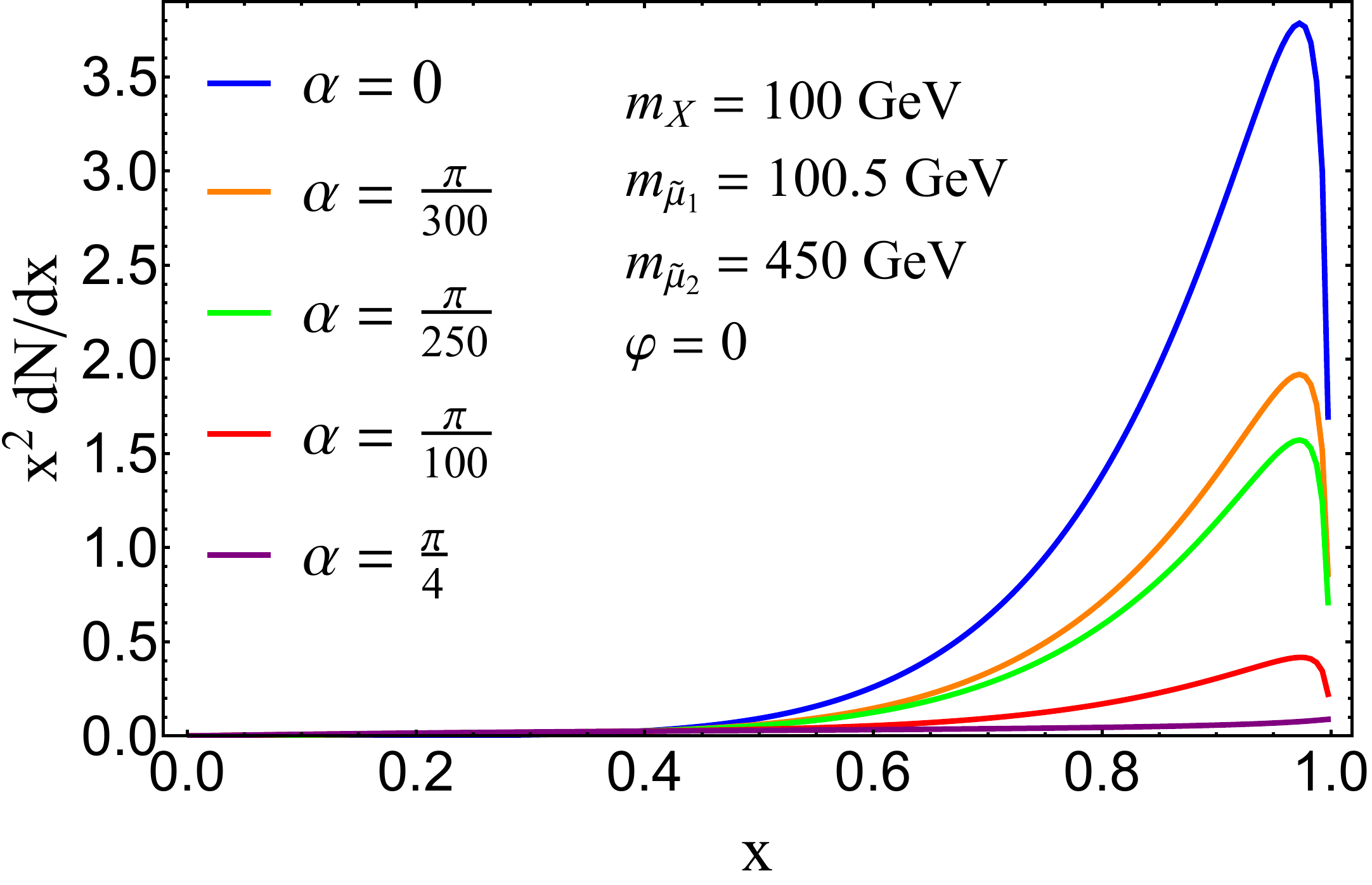}
    \vphantom{\includegraphics[width=0.47\textwidth,valign=m]{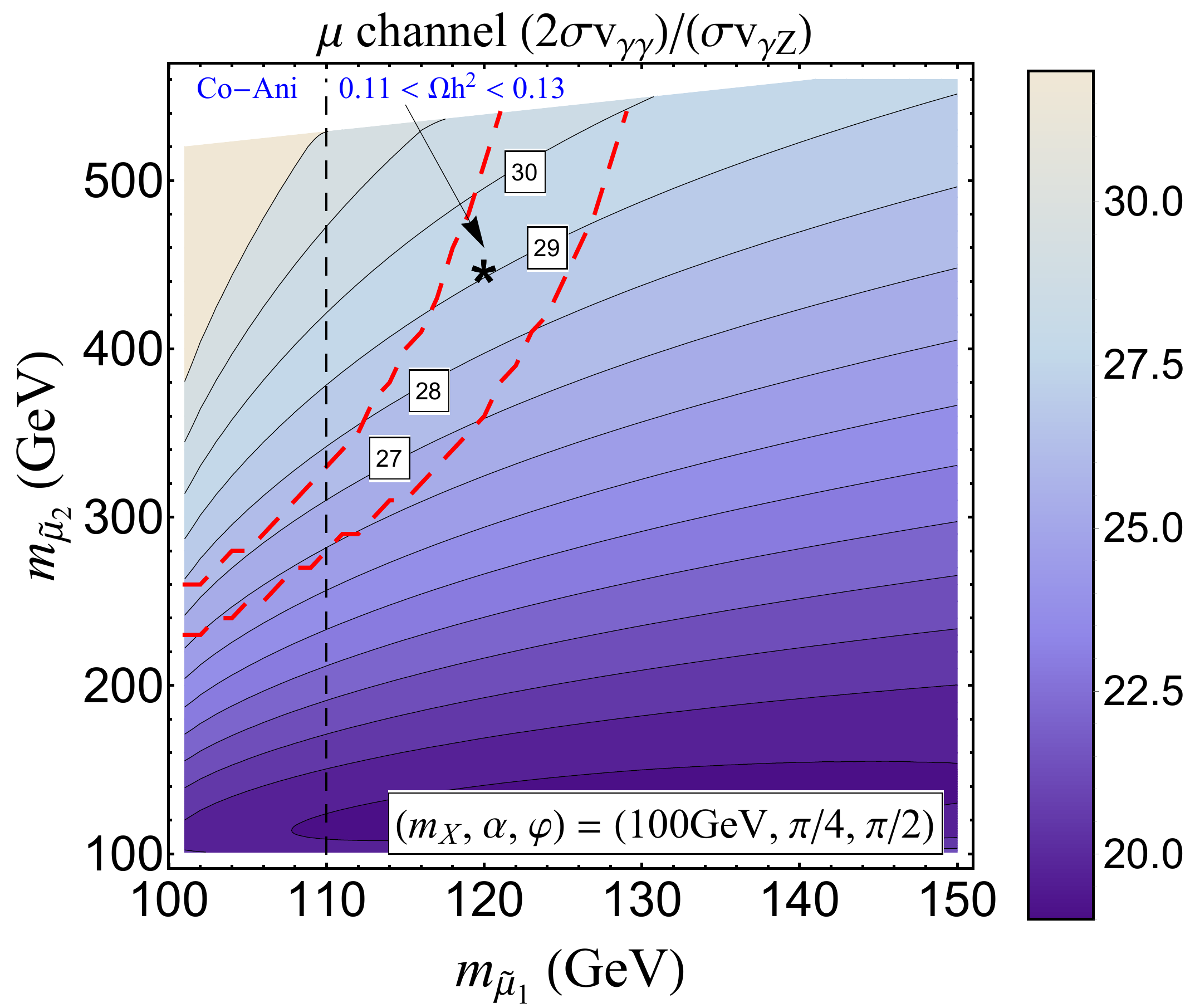}}
  }
  \subfloat[Ratio]{%
    \includegraphics[width=0.47\textwidth,valign=m]{Mu_ratio_100.pdf}%
  }
  \caption{\label{fig:gammaray}(a) The dependence of the IB spectrum on the mixing angle $\alpha$. (b) The $\gamma\gamma$ and $\gamma Z$ flux ratio at the MSSM couplings. The dashed contour encloses the region that reproduces the correct thermal relic density. }
\end{figure}

We first show the results of indirect detections. We are most interested in the gamma ray signals from dark matter annihilation, which are potentially observable at the galactic center and dwarf galaxies. These signals suffer less uncertainties during propagation compared with the charged cosmic rays so that they can draw more reliable limits to our parameter space. In our model \eqref{eq:lag}, the gamma ray signals are contributed by the internal bremsstrahlung (IB) $XX\rightarrow f\bar{f}\gamma$ and the one-loop suppressed annihilation $XX\rightarrow \gamma\gamma+\gamma Z$. If $\alpha=0$, the IB spectrum has a very sharp peak very close to the end point once $m_X$ and $m_1$ are sufficiently degenerate. This process is dominated by the photon emitted from the internal mediator. Nonzero mixing ($\alpha\neq 0$ and $\pi/2$), on the other hand, leads to a new $s$-wave amplitude, dominated by the photon emitted from the final state leptons. This amplitude is enhanced by soft and collinear photon emission and soon washes out the peaklike feature, as shown in the left panel of Fig.~\ref{fig:gammaray}. The mixing angle $\alpha$ and $CP$-violation phase $\varphi$ also have a significant impact on the flux ratio $2(\sigma v)_{\gamma\gamma}/(\sigma v)_{\gamma Z}$ of $\gamma\gamma$ and $\gamma Z$ line spectra. Observing these two lines is the smoking gun for a dark matter particle, and the flux ratio is then a direct probe of the model parameter space. In the right panel of Fig.~\ref{fig:gammaray}, this ratio is about $29$ at our benchmark $m_1=120\,\textrm{GeV}$ and $m_2=450\,\textrm{GeV}$. It is a very distinctive feature of our model at $\alpha=\pi/4$ and $\varphi=\pi/2$ since predictions from other models, for example CMSSM, are usually very different~\cite{Yaguna:2009cy}. Last but not least, the $CP$-violation phase $\varphi$ also gives rise to an interesting asymmetry in the photon helicities of the $\gamma\gamma$ final state, which can be as large as $20\%$ for $\tau$ channel. The origin of this asymmetry is a complex loop integral that has not appeared in the literature. We refer the interested readers to our work~\cite{Kumar:2016cum} for a detailed study on the gamma ray signals in the most general case.

\begin{figure}[t]
  \includegraphics[width=\textwidth]{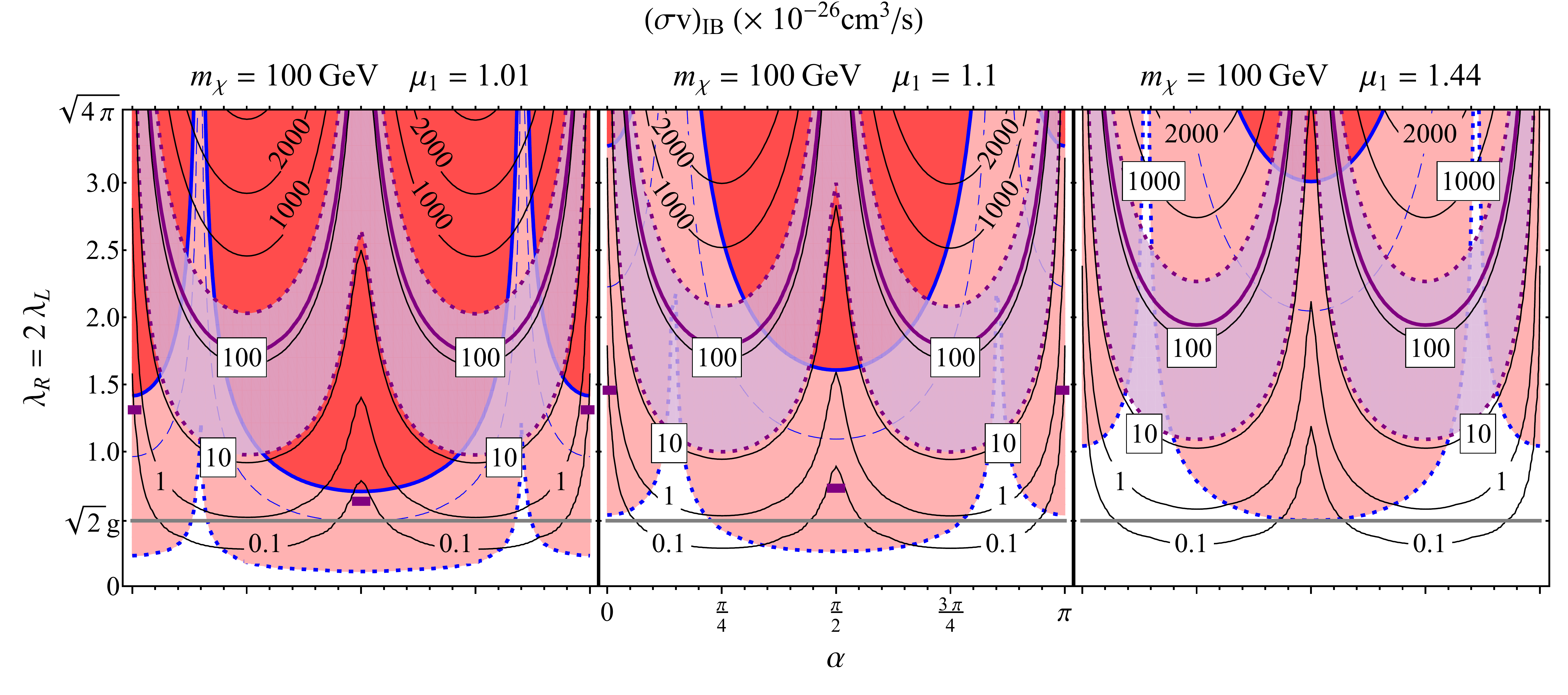}
  \caption{\label{fig:direct}Current and future constraints on the $(\lambda,\alpha)$ plane. The contours label the total IB cross section. The thick solid, thin dashed and thick dashed line corresponds respectively to LUX2013~\cite{Akerib:2015rjg}, LUX2016~\cite{Akerib:2016vxi} and LZ~\cite{Akerib:2015cja} projected limit. The purple band is the Fermi-LAT limit from continuum spectrum~\cite{GeringerSameth:2011iw}, while the thick purple lines at $\alpha=0$ and $\pi/2$ are the Fermi-LAT limit from line signal search~\cite{Ackermann:2015lka}. }
\end{figure}

Next, we investigate how current and future direct detections constrain our simplified model. Since Majorana dark matter does not couple to any colored mediators, the only way it interacts with nuclei is through the anapole moment at one loop. Usually one may think that this interaction is so weak that the signal region would be buried under the neutrino floor. However, this is not the case for $\mu$ channel. The anapole moment is inversely proportional to the lepton mass such that the smallness of the muon mass brings the interaction back to the target region of current and future direct detection experiments. In our work~\cite{Sandick:2016zut}, we calculate both the anapole moment and dark matter-nucleon interaction analytically, and then study the direct detection bounds on various aspects of our parameter space. The most interesting feature we found is a complimentarity between direct and indirect bounds. In Fig.~\ref{fig:direct}, we bring both bounds onto the same $(\lambda,\alpha)$ plane. At $\alpha=0$ and $\pi/2$ with a degenerate spectrum, where the Fermi-LAT line signal limits apply, the latest LUX limit is comparable to the line search one, while the future LZ will provide the most stringent limit, almost exhausting the entire parameter space, especially the SUSY case (the gray horizontal line in Fig.~\ref{fig:direct}). At nonzero mixing, LZ still outperforms the indirect detections for most cases, except for two ``blind spots'' around $\alpha=\pi/8$ and $7\pi/8$ where the anapole moment is severely suppressed. Then probing these regions relies heavily on indirect detections. 

Finally, we note that nonzero $\alpha$ and $\varphi$ can give rise to anomalous lepton dipole moments. It has been shown that in $\mu$ channel the $3\sigma$ deviation of the magnetic dipole moment from the Standard Model prediction can be fully accounted when $\varphi\sim\pi/2$~\cite{Fukushima:2014yia}. For $\tau$ channel, current experiments are too crude to probe our parameter space. However, following the spirit of simplified model, we do not view this as a hard constraint on our parameter space, since our simplified model is only supposed to capture the relevant information of how dark matter interacts with the visible sector. It could be another decoupled sector in the new physics that settles the dipole moments to the observed values.

\section{Discussion and conclusion}
One crucial component of our model is (at least) one light uncolored scalar mediator with mass $O(100\,\textrm{GeV})$, not too much heavier than the dark matter. Such a particle can escape the current LHC search, and the best equipment to identify it is future $e^{+}e^{-}$ colliders, like ILC~\cite{ILC} and CEPC~\cite{CEPC}. As a future project, we will study how to search such a particle at these colliders.

To conclude, we have studied possible direct and indirect detection signals of a general class of simplified model in which the Majorana dark matter only couples to a pair of uncolored scalar mediator and a Standard Model lepton. The mixing angle interpolates the peaklike and flat spectrum of the internal bremsstrahlung $XX\rightarrow f\bar{f}\gamma$. It also changes the flux ratio of $\gamma\gamma$ and $\gamma Z$ final state, making it distinctive from other models. The $CP$-violation phase gives rise to a polarization asymmetry in the $\gamma\gamma$ final state, which is an interesting feature for future experiments to explore. Future direct detections can probe most of parameter space, except for two ``blind spots'', which rely on indirect detections.

\newpage

\end{document}